\documentclass[preprint,amsmath,amssymb,aps]{revtex4}

\usepackage{epsfig,dsfont,amssymb,amsmath,amsthm,amsfonts,amsbsy,mathrsfs} 
\usepackage{graphicx} 
\usepackage{color} 
\usepackage{bm}
\usepackage{multirow} 

\usepackage{tikz}
\usetikzlibrary{arrows}

\newcommand{\BEQ}{\begin{equation}} 
\newcommand{\EEQ}{\end{equation}} 
\newcommand{\BEA}{\begin{eqnarray}} 
\newcommand{\EEA}{\end{eqnarray}}

\begin{document}

\title{Orthogonal run-and-tumble walks}

\author{L. Angelani$^{1,2}$}
\email{luca.angelani@roma1.infn.it}

\affiliation{$^1$ ISC-CNR, Institute for Complex Systems, P.le A. Moro 2, 00185 Rome, Italy}
\affiliation{$^2$ Dipartimento di Fisica, Sapienza Universit\`a di Roma, P.le A. Moro 2, 00185 Rome, Italy}



\begin{abstract}
Planar run-and-tumble walks with orthogonal directions of motion are considered. 
After formulating the problem with generic transition probabilities among the orientational states, we focus on the symmetric case, giving general expressions of the probability distribution function (in the Laplace-Fourier domain), the mean-square displacement and the effective diffusion constant in terms of transition rate parameters.
As case studies we treat and discuss 
two classes of motion, alternate/forward and isotropic/backward,
obtaining, when possible, analytic expressions of probability distribution functions in the space-time domain. 
We discuss at the end also the case of cyclic motion. 
Reduced (enhanced) effective diffusivity, with respect to the standard 2D active motion, is observed in the 
cyclic and backward (forward) cases.

\end{abstract}

\maketitle


\section{Introduction}

Random walk models with finite velocity describe many different physical and biological phenomena 
\cite{Weiss,Gold1951,Kac}, from the motion of electrons in metals \cite{Lor_1905} to the swimming of motile bacteria, such as {\it E.coli} \cite{Berg_book,Sch1993}. 
In many cases the particle motion can be described by the so called 
{\it run-and-tumble} models, in which the particle trajectory is a 
straight line interrupted by abrupt changes of motion direction.
Analytical results of run-and-tumble equations describing the time evolution of probability densities can be obtained 
in different simplified situations in one-dimensional space.
Higher dimensions are in general  quite harder to treat. Focusing on the two-dimensional case, an explicit expression of the probability density function
exists in the case of uniform turning-angle distribution \cite{Mar_2012}, while some other analytical results can be achieved under some approximations in the case of general turning-angle distribution \cite{Sev_2020}
or for Active Brownian particles \cite{Basu2018,Mala2020}.
Among planar motions, of some interest is the case of discrete turning-angle \cite{SBS_2020}, and, in particular, of orthogonal directions of motion 
\cite{godoy1997,Ors2000,Cin_2021}.

We investigate here the planar random motion of a particle which 
moves at constant speed along four different orthogonal directions of motion, switching between them at given rates.
The switching process is described by a transition probability matrix, whose elements are, in general, different one form each other.
While the problem has been previously treated in some special cases 
(see Ref. \cite{Cin_2021} and references within), 
we give here a 
very general and unified formulation, allowing us to obtain expressions valid for generic transition probabilities among the 
 different orientational states of the particle. We are then able to specialize the general formulae to various interesting case studies.
 
 The paper is organized as follows. 
 In Sec. II, we introduce the orthogonal run-and-tumble model,
 giving the formal general solution of the dynamical equations for the probability distribution functions. 
 In Sec. III, we analyze the symmetric case, reporting explicit expressions of the probability distribution function (in the  Laplace-Fourier domain) and the mean square displacement as a function of the transition rate parameters. We then specialize to some interesting case studies, such as 
 the alternate motion, with orthogonal switch of the direction of motion at each tumble event (as interesting byproduct we obtain the expression for the case of a 1D run-and-tumble particle with finite tumbling time),
 the isotropic motion, characterized by equal transition rates,
 and the cases of forward and backward motion, which turn out to be equivalent to the previous two cases with rescaled parameters.  
In Sec. IV we consider instead the case of a cyclic motion, where the particle’s direction of motion, after a tumble, rotates 90 degrees counterclockwise
Conclusions are drawn in Sec. V.

\section{Orthogonal run-and-tumble model}

We consider a  run-and-tumble particle in a plane which can move along two orthogonal directions of motion, parallel
to ${\hat{\bf x}}$ and ${\hat{\bf y}}$ axes.
Therefore there are only four possible self-propelling orientations for the particle,
i.e., 
$+\hat{\bf x}$ ($R$, Right),
$-\hat{\bf x}$ ($L$, Left),
$+\hat{\bf y}$ ($U$, Up),
$-\hat{\bf y}$ ($D$, Down).
We denote  with $P_\mu(x,t)$ -- with $\mu \in \{R,L,U,D\}$ --
the probability density functions (PDF) for the $\mu$-oriented particles.
Reorientation of the particle motion is described by a Poisson process with rate $\alpha$ 
and we denote with $\gamma_{_{\mu \nu}}$ 
the transition probability from state $\nu$ to state $\mu$: $\nu \to \mu$.
We note that in general $\gamma_{_{\mu \mu}}$ can be different  from zero, i.e. after a tumble the new direction chosen could be the same as the previous one.
The orthogonal run-and-tumble motion is
described by the following general equations for the PDFs
\begin{eqnarray}
\label{eq_r}
\frac{\partial P_{_R}}{\partial t} &= -& v \frac{\partial P_{_R} }{\partial x} 
- \alpha  P_{_R} + \alpha \sum_{\mu} \gamma_{_{R \mu}}\ P_{\mu} \\  
\label{eq_l}
\frac{\partial P_{_L}}{\partial t} &=&   v \frac{\partial P_{_L} }{\partial x} 
- \alpha  P_{_L} + \alpha \sum_{\mu} \gamma_{_{L \mu}}\ P_{\mu} \\  
\label{eq_u}
\frac{\partial P_{_U}}{\partial t} &= -& v \frac{\partial P_{_U} }{\partial y} 
- \alpha  P_{_U} + \alpha \sum_{\mu} \gamma_{_{U \mu}}\ P_{\mu} \\  
\label{eq_d}
\frac{\partial P_{_D}}{\partial t} &=&   v \frac{\partial P_{_D} }{\partial y} 
- \alpha  P_{_D} + \alpha \sum_{\mu} \gamma_{_{D \mu}}\ P_{\mu} 
\end{eqnarray}

We define the  vector ${\mathbf P}$
\begin{equation}
{\mathbf P} = 
\left(
\begin{array}{c}
P_{_R} \\
P_{_L} \\
P_{_U} \\
P_{_D}
\end{array}
\right)
\end{equation}
the derivatives matrix
\begin{equation}
{\mathbf D} = 
\left(
\begin{array}{cccc}
v \partial_x & 0 & 0 & 0  \\
0 & - v \partial_x & 0 & 0  \\
0 & 0 & v \partial_y & 0  \\
0 & 0 & 0 & - v \partial_y  
\end{array}
\right)
\end{equation}
and the transition matrix
\begin{equation}
\label{TMgen}
{\mathbf \Gamma} = 
\left(
\begin{array}{cccc}
\gamma_{_{R R}} & \gamma_{_{R L}} & \gamma_{_{R U}} &  \gamma_{_{R D}} \\
\gamma_{_{L R}} & \gamma_{_{L L}} & \gamma_{_{L U}} &  \gamma_{_{L D}} \\
\gamma_{_{U R}} & \gamma_{_{U L}} & \gamma_{_{U U}} &  \gamma_{_{U D}} \\
\gamma_{_{D R}} & \gamma_{_{D L}} & \gamma_{_{D U}} &  \gamma_{_{D D}} 
\end{array}
\right)
\end{equation}
with the constraint (probability conservation)
\begin{equation}
\sum_{\mu} \gamma_{_{\mu \nu}} = 1 
\end{equation}
We can write the Eq.s(\ref{eq_r}-\ref{eq_d}) in a concise form 
\begin{equation}
\label{ORTeq}
\frac{\partial {\mathbf P}}{\partial t} = 
- [ {\mathbf D} + \alpha (\mathds{1} - {\mathbf \Gamma}) ] {\mathbf P} 
\end{equation}
where $\mathds{1}$ is the identity matrix.
It is more convenient to work in the Laplace-Fourier domain
\begin{equation}
\widehat{\widetilde P} ({\bf k},s) = \int_0^\infty dt \ e^{-st}
\ \int d{\bf r} \ e^{i {\bf k}\cdot {\bf r}}
\ P({\bf r}, t) 
\end{equation}
where the symbols $\widetilde{\cdot}$ and $\widehat{\cdot}$ denote, respectively, Laplace and Fourier transforms. 
Eq.(\ref{ORTeq}) becomes
\begin{equation}
\label{FLORTeq}
[(s+\alpha) \mathds{1}  + {\mathbf D'} - \alpha {\mathbf \Gamma} ] {\widehat{\widetilde{\mathbf P}}} = 
 {\widehat{\mathbf P}}_0
\end{equation}
where the RHS is the Fourier transform of the initial distribution 
${\mathbf P}_0({\bf r}) = {\mathbf P}({\bf r},t\!=\!0)$
and the matrix $ {\mathbf D'} $ is 
\begin{equation}
{\mathbf D'} = 
\left(
\begin{array}{cccc}
-i v k_x   & 0 & 0 & 0  \\
0 & i v k_x & 0 & 0  \\
0 & 0 & -i v k_y & 0  \\
0 & 0 & 0 & i v k_y  
\end{array}
\right)
\end{equation}
By defining the matrix ${\mathbf A}$
\begin{equation}
\label{A_matr}
{\mathbf A} = (s+\alpha) \mathds{1}  + {\mathbf D'} - \alpha {\mathbf \Gamma}
\end{equation}
the Eq.(\ref{FLORTeq}) can be concisely written as
\begin{equation}
\label{FLORTeq2}
{\mathbf A} \ {\widehat{\widetilde{\mathbf P}}} = {\widehat{\mathbf P}}_0
\end{equation}
Then, the formal expression of the Laplace-Fourier transformed PDF, for generic initial conditions, can be written as
\begin{equation}
\label{PORT}
{\widehat{\widetilde{\mathbf P}}} = {\mathbf A}^{-1} \ {\widehat{\mathbf P}}_0
\end{equation}
\noindent 
Let us now specialize to the case of isotropic initial conditions
\begin{equation}
P_\mu ({\bf r},t=0) = \frac14 \delta({\bf r})   \hspace{2cm} \forall \mu \in \{ R,L,U,D \}
\end{equation}
corresponding in the Fourier space to
\begin{equation}
({\widehat{\mathbf P}}_0)_{\mu} = 
\frac14    \hspace{2cm} \forall \mu \in \{ R,L,U,D \}
\end{equation}
We are interested in the total distribution function, independent of particle orientation
\begin{equation}
P = P_{_R} +  P_{_L} +  P_{_U} +  P_{_D} 
\end{equation}
which can be then written, from Eq.(\ref{PORT})
\begin{equation}
{\widehat{\widetilde P}} = \frac14 \sum_{\mu, \nu} \ ({\mathbf A}^{-1})_{_{\mu \nu}} 
\end{equation}
The above expression allows us to obtain the probability distribution function as a sum
of the elements of the inverse of the matrix ${\mathbf A}$ defined in (\ref{A_matr}).
The obtained expression is very general, valid for generic transition probabilities $\gamma_{\mu \nu}$.

\noindent 
In the following section we give generic explicit solutions for the symmetric case, considering rotational symmetry and equivalence among the orientational states $R,L,U,D$. The general obtained expressions 
will allow us to specialize to few interesting case studies.
We analyze the case of isotropic transition probabilities, i.e.,
after a tumble the particle can assume with equal probability each one of the four possible propelling directions.
Another case we consider is the one with right reorientational angles, i.e., the particle orientation  switches between the two orthogonal directions ${\hat{\mathbf x}}$ and ${\hat{\mathbf y}}$.
An interesting byproduct of this planar motion is obtained projecting the solution onto the $x$ axis, resulting in a one dimensional motion with 3 states, considering a finite rest time during tumble events.
We also analyze the problems of forward and backward moving particle, which, after a tumble, can only move forward/backward or orthogonal to the previous direction of motion.

In a final section we analyze the case of cyclic motion, considering unidirectional rotational motion of the self-propelled direction.


\section{Symmetric case}

We consider here the symmetric case, where all the orientational states are equivalent and, moreover, symmetric rotational symmetry is assumed, considering equal transition probabilities from a given state towards the two perpendicular directions.
We can write the transition matrix as follow:
\begin{equation}
{\mathbf \Gamma} = 
\left(
\begin{array}{cccc}
\gamma_{_F} & \gamma_{_B} & \gamma_{_P} &  \gamma_{_P} \\
\gamma_{_B} & \gamma_{_F} & \gamma_{_P} &  \gamma_{_P} \\
\gamma_{_P} & \gamma_{_P} & \gamma_{_F} &  \gamma_{_B} \\
\gamma_{_P} & \gamma_{_P} & \gamma_{_B} &  \gamma_{_F} 
\end{array}
\right)
\label{symatri}
\end{equation}
where $\gamma_{_F}$, $\gamma_{_B}$ and $\gamma_{_P}$ are, respectively, forward, backward and perpendicular transition probabilities after a tumble event, satisfying the constraint
\begin{equation}
\gamma_{_F} + \gamma_{_B} + 2 \gamma_{_P}= 1 
\end{equation}
While in principle one can treat the symmetric cases by reabsorbing the diagonal terms of the transition matrix  in a rescaled tumbling rate, we prefer to maintain the original formulation with the presence of forward transition terms, allowing for generalisations to non-symmetric  cases, as for example in the presence of orientational dependent forward transition rates.\\
\noindent 
The simplified transition matrix with a reduced number of independent elements, with respect to the general case in Eq. (\ref{TMgen}), allows us to obtain explicit solutions of dynamical equations in a simple form.
By solving the linear equations (\ref{FLORTeq2}) for $P_{\mu}$
 -- or inverting the ${\bf A}$ matrix (\ref{A_matr}) and using (\ref{PORT}) --
we can obtain, after some algebra, the general expression of the PDF in the Laplace-Fourier space $(s,\bf{k})$ as a function of transition probability parameters:
\begin{equation}
{\widehat{\widetilde{P}}} = 
\frac{(s+\alpha_1)[(s+\alpha_1) (s+2\alpha_2) + k^2 v^2/2]}
{[k_x^2v^2+(s+\alpha_1)(s+\alpha_2)][k_y^2v^2+(s+\alpha_1)(s+\alpha_2)]-\alpha_2^2(s+\alpha_1)^2}
\label{PDF_sym}
\end{equation}
where $k^2=k_x^2+k_y^2$ and 
\begin{eqnarray}
\alpha_1 &= \alpha ( 1 + \gamma_{_B} - \gamma_{_F}) &=
2 \alpha (\gamma_{_B} + \gamma_{_P}) \\
\alpha_2 &= \alpha ( 1 - \gamma_{_B} - \gamma_{_F}) &= 
2 \alpha \gamma_{_P}
\end{eqnarray}
\noindent 
An interesting quantity characterizing the motion is the mean square displacement (MSD), obtained through the relation 
\cite{ksbook}
\begin{equation} 
r^2 = - \left. \nabla^2_{\bf k} \widehat{P} \right|_{{\bf k}\!=\!0}
\label{r2P}
\end{equation}
By deriving Eq.(\ref{PDF_sym}) we obtain the Laplace-transformed MSD
\begin{equation}
{\widetilde {r^2}} (s) = 
\frac{2v^2}{s^2} \frac{1}{s+\alpha_1} =
\frac{2v^2}{\alpha_1^2} \left[ \frac{\alpha_1}{s^2} - \frac1s +\frac{1}{s+\alpha_1} \right]
\label{r2sactive}
\end{equation}
corresponding, in the time domain, to
\begin{equation}
r^2(t) = \frac{2 v^2}{\alpha_1^2} \ \left[ \alpha_1 t - 1 + e^{-\alpha_1 t}\right]
\label{r2active}
\end{equation}
\noindent 
We note that this expression corresponds to the usual MSD for active particles with rescaled tumbling rate $\alpha_1$ 
\cite{Weiss,Mar_2012}.
It is worth also noting that a similar expression 
as (\ref{r2sactive}) can be obtained as special case of a more general form with continuous distribution of tumbling angles (see Eq. 45 of Ref.\cite{Detch} specialized to Poissonian tumbling).
The diffusive limit is obtained for $v,\alpha \to \infty$ with $v^2/\alpha$ constant \cite{Weiss}.
In this limit the PDF (\ref{PDF_sym}) reduces to the well known expression for the Brownian motion 
\begin{equation}
{\widehat{\widetilde{P}}}_{Diff} = \frac{1}{s + D k^2}
\end{equation}
corresponding to the time dependence
\begin{equation}
{\widehat{P}}_{Diff} = \exp{(-D k^2 t)}
\end{equation}
with $D$ the effective diffusion constant of the run-and-tumble particle 
\begin{equation}
D = \frac{v^2}{2\alpha_1} = \frac{v^2}{2\alpha}\  \frac{1}{2(\gamma_{_B}+ \gamma_{_P})}
\label{Deff_sym}
\end{equation}
Then, in the diffusive limit, the MSD reduces to the usual linear form
\begin{equation}
r^2(t) = 4 D t
\end{equation}

\noindent 
In the following subsections we specialize to some interesting case studies, reporting expressions of the PDFs and mean-square displacements and summarizing the values of effective diffusivity in Table I. 

\begin{table}
\label{tab1}
\begin{tabular}{ ||c|c|c|c|| } 
\hline
Model & Effective Diffusivity\\ 
\ & (in unit of $D_0 = v^2/2\alpha$) \\
\hline \hline
 Isotropic & $1$ \\ 
\hline
Alternate & $1$ \\
\hline
Backward  & $3/4$ \\
\hline
Forward  & $3/2$ \\
\hline
Cyclic  & $1/2$ \\
\hline

\end{tabular}
\caption{Long time effective diffusivity (in unit of standard run-and-tumble diffusivity in two dimensions $D_0=v^2/2\alpha)$ for the different analyzed run-and-tumble models.}
\end{table}


\subsection{Alternate and forward motions}

The first class of models we analyze is that including alternate and forward motions. We first consider the alternate motion, which has been quite extensively investigated in the past and then it serves as a benchmark of our results.

\subsubsection{Alternate motion}

We consider the case of a particle performing alternate motion along $\hat{\mathbf x}$ and $\hat{\mathbf y}$ axes. At each tumbling event the particle can switch between the two orthogonal directions of motion, as described in the following picture
\begin{center}
\begin{tikzpicture}
\tikzstyle{every node}=[font=\large,black]
\tikzset{vertex/.style = {shape=circle,draw,minimum size=1.5em}}
\tikzset{edge/.style = {->, >=stealth', line width=0.7pt}}
\node[vertex] (1) at  (2,0) {$U$};
\node[vertex] (2) at  (4,-2) {$R$};
\node[vertex] (3) at  (2,-4) {$D$};
\node[vertex] (4) at  (0,-2) {$L$};

\draw[edge] (1) to (2);
\draw[edge] (1) to (4);
\draw[edge] (2) to (1);
\draw[edge] (2) to (3);
\draw[edge] (3) to (2);
\draw[edge] (3) to (4);
\draw[edge] (4) to (1);
\draw[edge] (4) to (3);
\end{tikzpicture}
\end{center}
The transition matrix now reads
\begin{equation}
{\mathbf \Gamma} = \frac12
\left(
\begin{array}{cccc}
0 & 0 & 1 &  1 \\
0 & 0 & 1 & 1 \\
1 & 1 & 0 & 0 \\
1 & 1 & 0 & 0 
\end{array}
\right)
\end{equation}
corresponding to $\gamma_{_F}=\gamma_{_B}=0$ and $\gamma_{_P}=1/2$.
The PDF in this case is given by the following expression 
($\alpha_1=\alpha_2=\alpha$)
\begin{equation}
\label{PDF_AORT}
{\widehat{\widetilde{P}}} = 
\frac{(s+\alpha)[(s+\alpha)(s+2\alpha) + k^2 v^2/2]}
{[k_x^2v^2+(s+\alpha)^2][k_y^2v^2+(s+\alpha)^2]-\alpha^2(s+\alpha)^2}
\end{equation}
The mean square displacement is given by the usual form 
\begin{equation}
r^2(t) = \frac{2 v^2}{\alpha^2} \ \left[ \alpha t - 1 + e^{-\alpha t}\right]
\end{equation}
and the effective diffusivity in the diffusive limit reads
\begin{equation}
D = \frac{v^2}{2\alpha}
\end{equation}

The present case is of particular relevance, as we can write an explicit expression of the PDF in the real space. Indeed, we observe that the orthogonal motion in the $(x,y)$ coordinates reference corresponds to a sum of two independent run-and-tumble motions (with rescaled velocity $v/\sqrt{2}$ and tumbling rate $\alpha$) in the $\pi/4$ rotated coordinates reference $(x',y')$, 
as also observed in 
Ref.s \cite{Cin_2021,Smi2022}).
We can then write the explicit solution as a product
\begin{equation}
P(x,y,t;\alpha,v) = 
P_{1d}^0 \left(\frac{x+y}{\sqrt{2}}, t;\alpha,\frac{v}{\sqrt{2}}\right) \ 
P_{1d}^0 \left(\frac{x-y}{\sqrt{2}}, t;\alpha,\frac{v}{\sqrt{2}}\right)
\end{equation}
where $P_{1d}^0$ is the PDF of the $1D$ standard run-and-tumble motion \cite{Weiss,Mar_2012}
\begin{eqnarray}
P_{1d}^0(x,t;\alpha,v) & = & \frac{e^{-\alpha t/2}}{2} 
\bigg\{ 
\delta(x-v t) + \delta(x+v t)  \nonumber \\
 & +& 
\left[
\frac{\alpha}{2v}\ I_0\left(\frac{\alpha \Delta(x,t)}{2v}\right) +
\frac{\alpha t}{2 \Delta(x,t)}\  I_1\left(\frac{\alpha \Delta(x,t)}{2v}\right)  \right] \theta(v t-|x|)
\bigg\}
\end{eqnarray}
where we have explicitly indicated the parametric dependence on tumbling rate $\alpha$ and velocity $v$.
$I_0$, $I_1$  are the modified Bessel functions of zero and first order and
\begin{equation}
\Delta=\sqrt{v^2t^2- x^2}
\end{equation}
The reported results for the case of alternate motion can be used to obtain the exact solution of the one-dimensional run-and-tumble motion with finite values of the tumbling times. We provide this derivation in Appendix A.

\subsubsection{Forward motion}

Strictly related to the previous case is that of forward motion. 
In this case a particle, after a tumble, cannot move backward, but only forward or orthogonal to the previous direction of motion, as shown in the following diagram
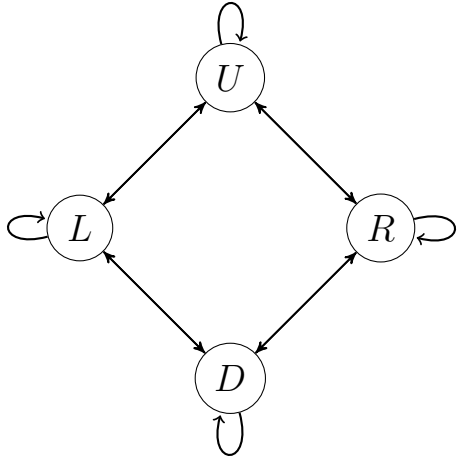
\begin{center}
\begin{tikzpicture}
\tikzstyle{every node}=[font=\large,black]
\tikzset{vertex/.style = {shape=circle,draw,minimum size=1.5em}}
\tikzset{edge/.style = {->, >=stealth', line width=0.7pt}}
\node[vertex] (1) at  (2,0) {$U$};
\node[vertex] (2) at  (4,-2) {$R$};
\node[vertex] (3) at  (2,-4) {$D$};
\node[vertex] (4) at  (0,-2) {$L$};
\draw[thick]  (1) edge [loop above] (1);
\draw[thick]  (2) edge [loop right] (2);
\draw[thick] (3) edge [loop below] (3);
\draw[thick] (4) edge [loop left] (4);
\draw[edge] (1) to (2);
\draw[edge] (1) to (4);
\draw[edge] (2) to (1);
\draw[edge] (2) to (3);
\draw[edge] (3) to (2);
\draw[edge] (3) to (4);
\draw[edge] (4) to (1);
\draw[edge] (4) to (3);
\end{tikzpicture}
\end{center}
The transition matrix is then of the form
\begin{equation}
{\mathbf \Gamma} = \frac{1}{3}
\left(
\begin{array}{cccc}
1 & 0 & 1 &  1 \\
0 & 1 & 1 & 1 \\
1 & 1 & 1 & 0 \\
1 & 1 & 0 & 1 
\end{array}
\right)
\end{equation}
corresponding to $\gamma_{_F}=\gamma_{_P}=1/3$ and $\gamma_{_B}=0$.
The PDF can be written as
\begin{equation}
{\widehat{\widetilde{P}}} = 
\frac{(s+\alpha_1)[(s+\alpha_1) (s+2\alpha_1) + k^2 v^2/2]}
{[k_x^2v^2+(s+\alpha_1)^2][k_y^2v^2+(s+\alpha_1)^2]-\alpha_1^2(s+\alpha_1)^2}
\end{equation}
where the effective tumbling rate is
\begin{equation}
\alpha_1 = \frac23 \alpha
\end{equation}
In other wards, the motion is the same of the alternate orthogonal case Eq.(\ref{PDF_AORT}), with the effective reduced tumbling rate $\alpha_1$,
The mean square displacement has the usual form, with the rescaled tumbling parameter $\alpha_1$
\begin{equation}
r^2(t) = \frac{2 v^2}{\alpha_1^2} \ \left[ \alpha_1 t - 1 + e^{-\alpha_1 t}\right]
\end{equation}
In the diffusive limit, contrary to the previous case, we observe an enhanced
diffusion of a factor $3/2$ 
\begin{equation}
D = \frac32 \ \frac{v^2}{2\alpha}
\end{equation}


\subsection{Isotropic and backward motions}
The second class of motions we consider is that including isotropic and backward cases. 

\subsubsection{Isotropic motion}
The isotropic motion refers to a particle that, after a tumble, chooses the new direction of motion among the four allowed ones in a isotropic way. 
The allowed transitions among states can be represented by the following schematic picture
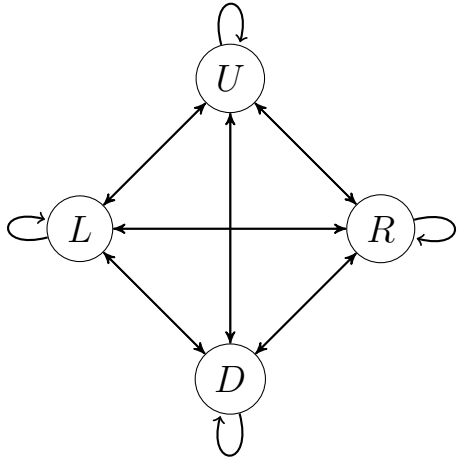
\begin{center}
\begin{tikzpicture}
\tikzstyle{every node}=[font=\large,black]
\tikzset{vertex/.style = {shape=circle,draw,minimum size=1.5em}}
\tikzset{edge/.style = {->, >=stealth', line width=0.7pt}}
\node[vertex] (1) at  (2,0) {$U$};
\node[vertex] (2) at  (4,-2) {$R$};
\node[vertex] (3) at  (2,-4) {$D$};
\node[vertex] (4) at  (0,-2) {$L$};
\draw[thick]  (1) edge [loop above] (1);
\draw[thick]  (2) edge [loop right] (2);
\draw[thick] (3) edge [loop below] (3);
\draw[thick] (4) edge [loop left] (4);
\draw[edge] (1) to (2);
\draw[edge] (1) to (3);
\draw[edge] (1) to (4);
\draw[edge] (2) to (1);
\draw[edge] (2) to (3);
\draw[edge] (2) to (4);
\draw[edge] (3) to (1);
\draw[edge] (3) to (2);
\draw[edge] (3) to (4);
\draw[edge] (4) to (1);
\draw[edge] (4) to (2);
\draw[edge] (4) to (3);
\end{tikzpicture}
\end{center}
The transition probabilities are all equal
\begin{equation}
\gamma_{\mu \nu} = \frac14
\hspace{2cm} \forall \mu,\nu \in \{ R,L,U,D \}
\end{equation}
and the transition matrix reads
\begin{equation}
{\mathbf \Gamma} = \frac14
\left(
\begin{array}{cccc}
1 & 1 & 1 &  1 \\
1 & 1 & 1 & 1 \\
1 & 1 & 1 & 1 \\
1 & 1 & 1 & 1 
\end{array}
\right)
\end{equation}
This case is then obtained from the previously derived formulae by setting $\gamma_{_F} = \gamma_{_B}=\gamma_{_P}=1/4$.
From the general expression (\ref{PDF_sym}), 
with $\alpha_1=\alpha$ and $\alpha_2=\alpha/2$, we have
\begin{equation}
{\widehat{\widetilde{P}}} = 
\frac{(s+\alpha)[(s+\alpha)^2 + k^2 v^2/2]}
{[k_x^2v^2+(s+\alpha)(s+\alpha/2)][k_y^2v^2+(s+\alpha)(s+\alpha/2)]-\alpha^2(s+\alpha)^2/4}
\label{PDF_iso}
\end{equation}
where $k^2=k_x^2+k_y^2$.

\noindent 
The mean square displacement now reads
\begin{equation}
r^2(t) = \frac{2 v^2}{\alpha^2} \ \left[ \alpha t - 1 + e^{-\alpha t}\right]
\end{equation}
and the diffusion constant is 
\begin{equation}
D = \frac{v^2}{2\alpha}
\end{equation}


\subsubsection{Backward motion}

The backward motion belongs to the same class of the previous case.
Here the particle, after a tumble, can only move backward or orthogonal to the previous direction of motion \cite{KO1999}.

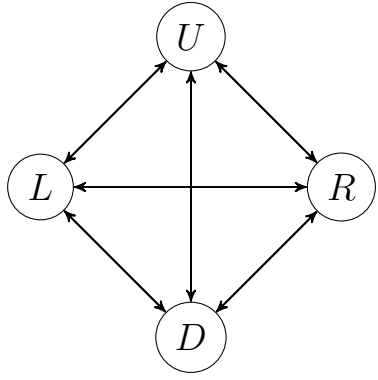
\begin{center}
\begin{tikzpicture}
\tikzstyle{every node}=[font=\large,black]
\tikzset{vertex/.style = {shape=circle,draw,minimum size=1.5em}}
\tikzset{edge/.style = {->, >=stealth', line width=0.7pt}}
\node[vertex] (1) at  (2,0) {$U$};
\node[vertex] (2) at  (4,-2) {$R$};
\node[vertex] (3) at  (2,-4) {$D$};
\node[vertex] (4) at  (0,-2) {$L$};
\draw[edge] (1) to (2);
\draw[edge] (1) to (3);
\draw[edge] (1) to (4);
\draw[edge] (2) to (1);
\draw[edge] (2) to (3);
\draw[edge] (2) to (4);
\draw[edge] (3) to (1);
\draw[edge] (3) to (2);
\draw[edge] (3) to (4);
\draw[edge] (4) to (1);
\draw[edge] (4) to (2);
\draw[edge] (4) to (3);
\end{tikzpicture}
\end{center}
The transition matrix reads
\begin{equation}
{\mathbf \Gamma} = \frac{1}{3}
\left(
\begin{array}{cccc}
0 & 1 & 1 &  1 \\
1 & 0 & 1 & 1 \\
1 & 1 &0 & 1 \\
1 & 1 & 1 & 0 
\end{array}
\right)
\end{equation}
corresponding to $\gamma_{_F}=0$ and $\gamma_{_B}=\gamma_{_P}=1/3$.
The PDF can be then written as
\begin{equation}
{\widehat{\widetilde{P}}} = 
\frac{(s+\alpha_1)[(s+\alpha_1)^2 + k^2 v^2/2]}
{[k_x^2v^2+(s+\alpha_1)(s+\alpha_1/2)][k_y^2v^2+(s+\alpha_1)(s+\alpha_1/2)]-\alpha_1^2(s+\alpha_1)^2/4}
\end{equation}
where $\alpha_1$ is the effective tumbling rate
\begin{equation}
\alpha_1 = \frac43 \alpha
\end{equation}
The above expression for the PDF is
similar to that of the isotropic case Eq.(\ref{PDF_iso}), indicating that the effect of not-forward motion is simply encoded in the rescaled tumbling rate $\alpha_1$.
The mean square displacement is then
\begin{equation}
r^2(t) = \frac{2 v^2}{\alpha_1^2} \ \left[ \alpha_1 t - 1 + e^{-\alpha_1 t}\right]
\end{equation}
In the diffusive limit, we observe a reduced diffusivity with respect to the isotropic case, with a diffusion constant reduced by a factor $3/4$:
\begin{equation}
D = \frac34 \ \frac{v^2}{2\alpha}
\end{equation}
We mention that in a recent work the backward model has been studied with generic transition rates, and the above expression for the MSD is recovered in the limit of equal rates \cite{Malli2022}.



\section{Cyclic Case}

Many interesting cases in nature show circular motion. This is for example the case of circular trajectories of bacteria close to surfaces \cite{LAU2006,DIL2011}. 
In our discrete orthogonal model this corresponds to consider a rotational cyclic motion \cite{Kol2004,OGZ2020}, with the following sequence of orientation switches
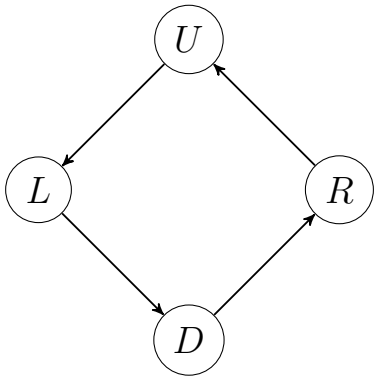
\begin{center}
\begin{tikzpicture}
\tikzstyle{every node}=[font=\large,black]
\tikzset{vertex/.style = {shape=circle,draw,minimum size=1.5em}}
\tikzset{edge/.style = {->, >=stealth', line width=0.7pt}}
\node[vertex] (1) at  (2,0) {$U$};
\node[vertex] (2) at  (4,-2) {$R$};
\node[vertex] (3) at  (2,-4) {$D$};
\node[vertex] (4) at  (0,-2) {$L$};
\draw[edge] (1) to (4);
\draw[edge] (2) to (1);
\draw[edge] (3) to (2);
\draw[edge] (4) to (3);
\end{tikzpicture}
\end{center}
Without loss of generality we are considering here the case of anti-clockwise motion.
The transition matrix is not expressed by the symmetric form (\ref{symatri}),
and it now reads
\begin{equation}
\label{gamma_cyc}
{\mathbf \Gamma} = 
\left(
\begin{array}{cccc}
0 & 0 & 0 &  1 \\
0 & 0 & 1 & 0 \\
1 & 0 & 0 & 0 \\
0 & 1 & 0 & 0 
\end{array}
\right)
\end{equation}
Also in this case it is possible to give an explicit expression of the PDF in the Laplace-Fourier space.
Proceeding as before, we can solve the linear equations (\ref{FLORTeq2}) for $P_{\mu}$ using the above expression 
of transition matrix (\ref{gamma_cyc}). We finally obtain:
\begin{equation}
{\widehat{\widetilde{P}}} = 
\frac{(s+2\alpha)[(s+\alpha)^2+\alpha^2]+(s+\alpha) k^2 v^2/2}
{[k_x^2v^2+(s+\alpha)^2][k_y^2v^2+(s+\alpha)^2]-\alpha^4}
\label{Pcyc}
\end{equation}

\noindent
From (\ref{r2P}) we can obtain the mean square displacement 
in the Laplace domain
\begin{equation}
{\widetilde {r^2}} (s) = 
\frac{2v^2}{s^2} \frac{s+\alpha}{(s+\alpha)^2+\alpha^2} =
\frac{v^2}{\alpha} \left[ \frac{1}{s^2} - 
\frac{1}{(s+\alpha)^2+\alpha^2} \right]
\end{equation}
corresponding, in the time domain, to
\begin{equation}
r^2(t) = \frac{v^2}{\alpha^2} \ \left[ \alpha t - e^{-\alpha t} \sin{(\alpha t)} \right]
\label{r2cyc}
\end{equation}
This expression differs from the previous one (\ref{r2active}). However, it has similar asymptotic behaviors: ballistic at short time $r^2 \simeq v^2 t^2$, and diffusive at long time, $r^2 \simeq v^2 t /\alpha$.
In the diffusive limit, $v,\alpha \to \infty$ with $v^2/\alpha$ constant,
the PDF (\ref{Pcyc}) reduces again to the well known expression
\begin{equation}
{\widehat{\widetilde{P}}}_{Diff} = \frac{1}{s + D k^2}
\end{equation}
with 
\begin{equation}
D = \frac12 \ \frac{v^2}{2\alpha}
\end{equation}
The cyclic rotational motion results in a slower diffusion of the particle and the effective diffusivity is reduced by a factor two with respect to the standard active motion (see Table I).

\section{Conclusions}
In this work we have treated the planar run-and-tumble walk with orthogonal directions of motion. After formulating the general problem with generic transition probabilities matrix, we focused on symmetric cases, giving analytic expressions of PDF, see Eq.(\ref{PDF_sym}),
mean-square displacements (\ref{r2active})
and effective diffusivity (\ref{Deff_sym})
in terms of rescaled tumbling rates. The obtained general formualae have been then specialized to some interesting cases previously investigated in the literature with different approaches (alternate/forward and isotropic/backward motions).
We finally discussed the case of circular cyclic motion, reporting expressions of PDF (\ref{Pcyc}) and MSD (\ref{r2cyc}).
The reported formulation allows us to treat in a simple way discrete orientational motions, making possible to extend the analysis to 
different and more complex situations, such as, for example,  
the cases of run-and-tumble walks with orientational dependent motilities or drift terms, with non-instantaneous tumble events, with non-orthogonal directions of motion, 
or also extending the analysis of some orthogonal models to higher dimensions or to stochastic resetting processes \cite{Smi2022,Eva2018,Mas2019}.

\section*{Acknolegements}
I acknowledge financial support from the MUR PRIN2020 project
2020PFCXPE.
I thank Roberto Garra for useful discussions and comments.

\section*{Appendix A. 1D Run-and-Tumble with finite tumbling time}
An interesting byproduct of the results reported in Sec.3A is obtained considering the marginal distribution, i.e., projecting the solution onto the
$\hat{\bf x}$ axis. 
It is easy to recognize that this corresponds to a one-dimensional run-and-tumble motion
with exponentially distributed rest time during tumbling (with the same mean value  $1/\alpha$ 
of the run-time). In other words we have a 1D three-states model in which the particle
alternates run motion and rest periods switching between them at the rate $\alpha$ 
\cite{SBS_2020,Ang_EPL_2013,Basu2020}.
By denoting with $P^{(3s)}_{1d}$ the PDF of this one-dimensional three-states motion, we have
\begin{equation}
{\widehat{\widetilde{P}}}^{(3s)}_{1d} (k,s) =  {\widehat{\widetilde{P}}} (k_x=k, k_y=0, s)
\end{equation}
where the RHS is the previously obtained quantity (\ref{PDF_AORT}).
We finally obtain
\begin{equation}
{\widehat{\widetilde{P}}}^{(3s)}_{1d} (k,s) =  
\frac{1}{2(s+\alpha)}\ 
\left[
\frac{(s+2\alpha)^2}{s(s+2\alpha)+k^2v^2}
+1
\right]
\end{equation}
in agreement with Eq.(17) of Ref. \cite{Ang_EPL_2013}
-- by setting in that reference
$\psi(t)  = \alpha e^{-\alpha t}$,
${\psi}(s) = \alpha/(s+\alpha)$,
$\tau_{_T}=1/\alpha$,
$P_0=(s+\alpha)/[(s+\alpha)^2+k^2v^2]$.
The explicit expression of the PDF in the $(x,t)$ domain can be 
obtained by performing the inverse Laplace-Fourier transform of the above expression, giving rise, after some algebra

\begin{eqnarray}
P^{(3s)}_{1d}(x,t) & = & \frac{e^{-\alpha t}}{4} \bigg\{
2 \delta (x) + \delta(x-vt) + \delta(x+vt)  \nonumber \\
 & +& \left.
\left[
\frac{2\alpha}{v}\ I_0\left(\frac{\alpha \Delta(x,t)}{v}\right) +
\frac{\alpha t}{\Delta(x,t)}\  I_1\left(\frac{\alpha \Delta(x,t)}{v}\right) + \right. \right. \nonumber \\
& + &\left. \left.
\frac{\alpha^2}{v}\ \int_{|x|/v}^{t} dt'\ I_0\left(\frac{\alpha \Delta(x,t')}{v}\right)
\right] \theta(vt-|x|)
\right\}
\end{eqnarray}
where $\Delta=\sqrt{v^2t^2- x^2}$ and $I_0$, $I_1$  are the modified Bessel functions of zero and first order.
We note that the above expression is in agreement with 
that reported in \cite{Basu2018} (expressed in a different form) and also with
the expression reported by Kolensik \cite{Kole2014} considering the sum of two independent telegraph processes on a line. One can recognize, indeed, that the three-states run-and-tumble process with tumbling rate $\alpha$ and velocity $v$ (run-right, run-left, tumble-rest) can be mapped into a process which is the sum of two independent two-states processes, each one with tumbling rate $\alpha/2$ and velocity $v/2$. Indeed, when the two processes correspond to run motions in the same direction one has twice the velocity of the single process, while when they correspond to two motions in opposite directions one has a rest situation.
Moreover, the rate at which happens a tumble of one of the two processes is twice the single rate. 


\end{document}